\documentclass{article}

\usepackage{arxiv}

\usepackage[utf8]{inputenc} % allow utf-8 input
\usepackage[T1]{fontenc}    % use 8-bit T1 fonts
\usepackage{hyperref}       % hyperlinks
\usepackage{url}            % simple URL typesetting
\usepackage{booktabs}       % professional-quality tables
\usepackage{amsfonts}       % blackboard math symbols
\usepackage{nicefrac}       % compact symbols for 1/2, etc.
\usepackage{microtype}      % microtypography
\usepackage{lipsum}

\usepackage[ruled,vlined]{algorithm2e}
\usepackage{graphicx}

\title{AVaN Pack: An Analytical/Numerical Solution for Variance-Based Sensitivity Analysis}

\author{
  Eduardo Vasconcelos\\
  Academic Departament of Electronic Systems Control\\
  Federal Institute of Pernambuco\\
  Recife, Pernambuco, Brazil \\
  \texttt{eduardo.vasconcelos@recife.ifpe.edu.br} \\
   \And
  Adriano Souza\\
  Academic Departament of Electronic Systems Control\\
  Federal Institute of Pernambuco\\
  Recife, Pernambuco, Brazil \\
  \texttt{adrianosouza@recife.ifpe.edu.br} \\
  \And
  Kelvin Dias\\
  Center of Informatic\\
  Federal University of Pernambuco\\
  Recife, Pernambuco, Brazil \\
  \texttt{kld@cin.ufpe.br} \\
}

\begin{document}
\maketitle

\begin{abstract}
Sensitivity analysis is an important concept to analyze the influences of parameters in a system, an equation or a collection of data. The methods used for sensitivity analysis are divided into deterministic and statistical techniques. Generally, deterministic techniques analyze fixed points of a model whilst stochastic techniques analyze a range of values. Deterministic methods fail in analyze the entire range of input values and stochastic methods generate outcomes with random errors. In this manuscript, we are interested in stochastic methods, mainly in variance-based techniques such as Variance and Sobol indices, since this class of techniques is largely used on literature. The objective of this manuscript is to present an analytical solution for variance based sensitive analysis. As a result of this research, two small programs were developed in Javascript named as AVaN Pack (\textbf{A}nalysis of \textbf{Va}riance  through \textbf{N}umerical solution). These programs allow users to find the contribution of each individual parameter in any function by means of a mathematical solution, instead of sampling-based ones.
\end{abstract}

% keywords can be removed

\section{Introduction}
The concept of sensitivity analysis is well known in the literature and it is used in several scientific areas to determine the impact of parameters in complex systems. There are a large number of techniques on literature, that can be divided into statistical and deterministic methods \cite{hendrickson}. Among these techniques, we can mention the Partial Differential Approach, the Rank of Correlation, the Regression method \cite{Allard2009} and Variance-Based Methods such as ANOVA (Analysis of Variance), and Sobol indices \cite{Archer1997}. The sensitivity analysis concept is widely used in many areas such as finance \cite{xu2005}, \cite{BORGONOVO2010}, \cite{Yoo2018}, engineering \cite{Oppio2017}, Biology \cite{Cogan2017} \cite{Gay2017} and many other areas.

Deterministic approaches consist in mathematically find the contribution of parameters in a function. So In this concept, the analysis is performed considering specific points over the range of inputs. Partial Derivative analysis \cite{Allard2009} is a deterministic technique that determine the contribution of parameters by analyzing the differential of parameter $x_i$ over the function $f(x_1,x_2,...,x_n)$ for specific values of $X$. The main advantage of this approach is that results do not present random errors. The main disadvantage, however, is that it fails in analyze the global scope of complex functions due to the fact that this technique's class analyzes deterministic points (local analysis). The main disadvantage, however, is that due to the fact that this class of technique analyzes deterministic points (local analysis) it fails in analyze the global scope of a complex function.

Statistic approaches, on the other hand, consist of creating a set of data series through the generation of random values to be passed as parameters to function $f(X)$. With the data series, it's possible to generate statistics such as the Correlation of Pearson between each parameter and the function output. The main advantage of this concept is that it can analyze the range of values of each parameter. The problem with this concept is due to its random nature, since it generates results with random error, what can decreases the overall quality of the analysis, as discussed in \cite{hendrickson}.

Besides the problems presented on both concepts of sensitivity analysis, for complex multi-variable functions $f(X)$, both concepts can fail to analyze certain regions of function $f$. For deterministic analysis, the person responsible for planning the analyzes can empirically choose values that cover only some regions of $f$. For statistic concept, the random nature of analyzes can bypass some regions of function $f$, even with a great number of samples.

To solve the problems presented, we propose a method to solve statistical sensitivity analysis of continuous functions through analytical solution. In this paper, our proposal focuses on the variance-based class of statistic techniques such as the indices of Sobol and Variance, since they are largely studied on literature \cite{Allard2009}\cite{Iooss2015}. 

As a result of this research, we have developed two scripts, coded in Javascript, that can find the contributions of parameters in a continuous function through of numerical solution. To this set of scripts, we have given the name of AVaN Pack (\textbf{A}nalysis of \textbf{Va}riance through \textbf{N}umerical solution).

This work is organized as following: section 2 presents the method used for analytically solve the Variance and the Sobol indices; section 3 presents and discusses a practical example commonly used on literature comparing the results produced by the proposed method with those obtained on \cite{Allard2009}; section 4 presents the scripts of AVaN Pack; and section 5 presents the considerations and future works.

\section{Analytically solving the indices of Variance and Sobol} \label{sec:models}

Let $f(X)$ be a function and $X={x_0,x_1,x_2,...,x_n} $ an array of input parameters. $S_i$ is the first level sensitivity contribution of parameter $x_i$ over $f(X)$ (named as "First Order index") that, in analysis of variance, can be obtained by [2]:
\begin{equation} \label{eq:1}
S_i=\frac{Var(f(x_i))}{Var(f(X))}
\end{equation}
where $Var(f(x_i))$ corresponds to the variance of function $f(X)$ by varying the parameter $x_i$ and keeping the other parameters unchanged, and $Var(f(X))$ is the variance of function $f(X)$.

The method used for calculating the Sobol index $So_i$ is different from $S_i$ since it considers the variance decomposition of $f(X)$ \cite{Archer1997}. So, the index $So_i$ is defined as:
\begin{equation} \label{eq:2}
So_i=\frac{V_i}{V(Y)}=\frac{Var_{x_i}(E_{-x_i}[f(X)|x_i])}{V(Y)}
\end{equation}
with
\begin{equation} \label{eq:3}
V(Y) = \sum_i^n{V_i}+\sum_{i,j}^n{V_{ij}}+\dots+V_{1,2,3,\dots,n}=Var_{x_i}(E_{-x_i}[f(X)|x_i])+E_{x_i}[Var_{-x_i}(f(X)|x_i)]
\end{equation}
\verb|Equations| \ref{eq:1} and \ref{eq:2} represent the first order indices of both analyses of variance and Sobol techniques. For the other orders we have:
\begin{equation} \label{eq:4}
S_{i..j}=\frac{Var(f(x_i...x_j))}{Var(f(X))}
\end{equation}
and
\begin{equation} \label{eq:5}
So_{i..j}=\frac{Var_{x_{i\dots j}}(E_{-x_{i\dots j}}[f(X)|x_{i\dots j}])-Var_{x_i}(E_{-x_i}[f(X)|x_{i}])-\dots-Var_{x_j}(E_{-x_j}[f(X)|x_{j}])}{V(Y)}
\end{equation}

In general, several authors have used sampling techniques to analyze functions using variance based methods in order to determine the variances of \verb|Equations| \ref{eq:1}, \ref{eq:2}, \ref{eq:4} and \ref{eq:5}. To solve the variance of any function analytically, the following expression can be used:

\begin{equation} \label{eq:6}
Var(f(x)) = E[f(x)^2]-E[f(x)]^2
\end{equation}
The expected value $E[f(x)]$ of a function can be obtained as:
\begin{equation} \label{eq:7}
E[f(x)]=\int{f(x)\varphi(f(x))}dx
\end{equation}
where $\varphi(x)$ is the Probability Distribution Function of $f(x)$. It is not trivial and some times impossible to find the Probability Distribution Function of complex equations. So, we can use the mathematical Expected Value of a function denoted by:
\begin{equation} \label{eq:8}
E[f(x)]=(b-a)^{-1}\int_a^b{f(x)}dx,b>a
\end{equation}
The problem of \verb|Equation| \ref{eq:8} is that to obtain the expected value of $f(x)$ is necessary a range of values. However, as all studies of the sensitivity analysis are performed under a range of values, the solution presented in \verb|Equation| \ref{eq:8} is feasible to our problem. So, to compute the term $E[f(x)^2]$, we can use the following expression:
\begin{equation} \label{eq:9}
E[f(x)^2]=(b-a)^{-1}\int_a^b{f(x)^2}dx,b>a
\end{equation}
Based on \verb|Equations| \ref{eq:6} to \ref{eq:9} we can find the variance of a function $f(x)$. But, these equations just calculate the first dimension variance, and to find the first order contribution we need the $n$ dimension variance of $f(X)$;  in other words, we need to find $Var(f(X))$. So, we can make use of the Fubini's Theorem for any integrable and continuous equation to find this answer:
\begin{equation} \label{eq:10}
\int_{y_{min}}^{y_{max}}{\int_{x_{min}}^{x_{max}}{f(x,y)}}dxdy=\int_{x_{min}}^{x_{max}}{\int_{y_{min}}^{y_{max}}{f(x,y)}}dydx
\end{equation}
So, from property presented in \verb|Equation| (10) we have:
\begin{equation} \label{eq:11}
E[f(X)]=(\prod_k^n{b_k-a_k})^{-1}\int_{a_1}^{b_1}\int_{a_2}^{b_2}
\int_{a_3}^{b_3}...\int_{a_n}^{b_n}{f(X)}dx_n...dx_3 dx_2 dx_1
\end{equation}
Thus, from \verb|Equations| \ref{eq:6}, \ref{eq:8}, \ref{eq:9} and \ref{eq:11} we can calculate the indices presented in \verb|Equations| \ref{eq:1}, \ref{eq:2}, \ref{eq:4} and \ref{eq:5}.

\subsection{The First Order Percentage Contribution} \label{sec:The First Order Percentage Contribution}

The problem of calculating $Var(f(x))$ and $V(Y)$ is the number of integrations that are required, since computationally, these values are generally only obtained through numerical solutions. But, to compute the first order indices for equation $f(X)$ we can consider these indices in percentage terms. So, we can write the percentage variance indices as:
\begin{equation} \label{eq:12}
S_i^\%=S_i\left (\sum_k^n S_k  \right )^{-1}
\end{equation}
Substituting the indices $S_i$ by \verb|Equation| \ref{eq:1} we have:
\begin{equation} \label{eq:13}
S_i^\%= \frac{Var(f(x_i))}{Var(f(X))} \left (\sum_k^n \frac{Var(f(x_k))}{Var(f(X))}  \right )^{-1}
\end{equation}
As $Var(f(X))$ is fixed and behaves as a constant in \verb|Equation| \ref{eq:13}, we have:
\begin{equation} \label{eq:14}
S_i^\%= \frac{Var(f(x_i))}{Var(f(X))} \left (\frac{1}{Var(f(X))}\sum_k^n Var(f(x_k))  \right )^{-1}
\end{equation}
Finally, the percentage index of variance is defined by \verb|Equation| \ref{eq:15}:
\begin{equation} \label{eq:15}
S_i^\%= Var(f(x_i)) \left (\sum_k^n Var(f(x_k))  \right )^{-1}
\end{equation}
Following the same idea for Sobol indices formula, we can derive the percentage contribution of Sobol method as follow:
\begin{eqnarray}\label{eqArr:1}
So_i^\% &=& So_i\left(\sum_k^n So_k\right)^{-1}\\
 &=& \frac{Var_{x_i}(E_{-x_i}[f(X)|x_i])}{V(Y)}\left(\sum_k^n \frac{Var_{x_k}(E_{-x_k}[f(X)|x_k])}{V(Y)} \right)^{-1}\\
 &=&\frac{Var_{x_i}(E_{-x_i}[f(X)|x_i])}{V(Y)}\left(\frac{1}{V(Y)} \sum_k^n Var_{x_k}(E_{-x_k}[f(X)|x_k]) \right)^{-1}\\
 &=&Var_{x_i}(E_{-x_i}[f(X)|x_i])\left(\sum_k^n Var_{x_k}(E_{-x_k}[f(X)|x_k]) \right)^{-1}
\end{eqnarray}

The \verb|Equations| \ref{eq:15} and \ref{eqArr:1} represent the first order contribution of parameters of $f(X)$ both for the variance analysis and for the Sobol analysis and can be efficiently calculated, since, we do not have to calculate $n$ integrals to find $Var(f(X))$. It's important to mention that $S_i^\%$ and $So_i^\%$ do not represent the index of variance neither the Sobol index of $f(X)$. It represents just the individual percentage contribution of parameter $x_i$ for variance and Sobol sensitivity techniques.

\section{Practical Example} \label{sec:Practical Example}

To demonstrate the use of the analytical solution proposed in this manuscript for both Variance and Sobol indices, we reproduce the experiment presented in \cite{Allard2009} by extracting the variance indices of Ishigami Function:
\begin{equation} \label{eq:16}
Y=sin(x)+7(sin(y))^2+0.1(z^4)sin(x)
\end{equation}

\subsection{Variance indices}\label{subsec:Example Variance}
To obtain the percentage first order variance of \verb|Equation| \ref{eq:16} we need to calculate the Expected Value presented in \verb|Equation| \ref{eq:6}, considering that the range used in \cite{Allard2009} is $(-\pi /10,\pi /10)$ for x, y and z. So, the integrations for functions $Var(f(x))$, $Var(f(y))$ and $Var(f(z))$ are:
\begin{equation} \label{eq:17}
Var(Y(x)) = E[Y(x)^2] - E[Y(x)]^2
\end{equation}
where:
\begin{eqnarray*}
E[Y(x)^2] &=& (\pi /10-(-\pi /10))^{-1}\int_{-\pi /10}^{\pi /10}{(sin(x)+7(sin(y))^2+0.1(z^4)sin(x))^2}dx \\
&=& \frac{4900\pi(cos(4y)-4cos(2y))-5\sqrt{10-2\sqrt{5}}(z^4+10)^2+4\pi(z^8+20z^4+2775)}{4000(\pi /10-(-\pi /10))}\\
\end{eqnarray*}
and
\begin{eqnarray*} 
E[Y(x)]^2 &=& \left ((\pi /10-(-\pi /10))^{-1}\int_{-\pi /10}^{\pi /10}{(sin(x)+7(sin(y))^2+0.1(z^4)sin(x))}dx \right)^2 \\
&=&\left(\frac{7\pi sin^2(y)}{5(\pi/10 - (-\pi/10))}\right)^2
\end{eqnarray*}
For the variances of $Y(y)$ and $Y(z)$ we have:
\begin{equation} \label{eq:18}
Var(Y(y)) = E[Y(y)^2] - E[Y(y)]^2
\end{equation}
\begin{eqnarray*}
E[Y(y)^2] &=& (\pi /10-(-\pi /10))^{-1}\int_{-\pi /10}^{\pi /10}{(sin(x)+7(sin(y))^2+0.1(z^4)sin(x))^2}dy \\
&=& 5\pi^{-1}(0.371384 + 0.0628319 z^4 + 
  0.00314159 z^8 + \\ &&(-0.314159 - 0.0628319 z^4 - 0.00314159 z^8) cos(    2x) + \\&& (0.283733 + 0.0283733 z^4) sin(x))\\
\end{eqnarray*}
\begin{eqnarray*}
E[Y(y)]^2 &=& \left((\pi /10-(-\pi /10))^{-1}\int_{-\pi /10}^{\pi /10}{(sin(x)+7(sin(y))^2+0.1(z^4)sin(x))}dy\right)^2 \\
&=& \pi^{-2}(25 ((0.0628319 z^4 + 0.628319) sin(x) + 0.141866))\\
\end{eqnarray*}
and
\begin{equation} \label{eq:19}
Var(Y(z)) = E[Y(z)^2] - E[Y(z)]^2
\end{equation}
\begin{eqnarray*}
E[Y(z)^2] &=& (\pi /10-(-\pi /10))^{-1}\int_{-\pi /10}^{\pi /10}{(sin(x)+7(sin(y))^2+0.1(z^4)sin(x))^2}dy \\
&=& (\pi/5)^{-1} \left( \frac{7\pi(500000+\pi^4)sin(x)sin^2(y)}{1.25\times 10^6} \right) + \\ && (\pi/5)^{-1}\left(\left[ \pi/5 + \pi^5/(1.25 \times 10^6) + (\pi^9/4.5 \times 10^{11}) \right] sin^2(x)+\frac{49 \pi sin^4(x)}{5} \right) \\
\end{eqnarray*}
\begin{eqnarray*}
E[Y(z)]^2 &=& \left((\pi /10-(-\pi /10))^{-1}\int_{-\pi /10}^{\pi /10}{(sin(x)+7(sin(y))^2+0.1(z^4)sin(x))}dz\right)^2 \\
&=& (\pi/5  (sin(x) + 7 sin^2(y)) + 0.000122408 sin(x))^2\\
\end{eqnarray*}

As we are interested in first-order influences, we need to define the values assumed to unchanged variables on equations, since in \cite{Allard2009} authors did not present these values explicitly. So, In this manuscript, we have fixed the values of x, y and z in 0, since it is the median value of the range $-\pi/10$ and $ \pi/10$

After solving \verb|Equations| \ref{eq:17}, \ref{eq:18} and \ref{eq:19}, we have obtained the following values: $Var(Y(x))=0.0322554$, $Var(Y(y))=0.0400963$ and $Var(Y(z))=0$. From the variances, we can obtain the percentage contribution using \verb|Equation| \ref{eq:15} obtaining the following results: $S_x^\%=44.58\%$, $S_y^\%=55.41\%$ and $S_z^\%=0\%$. 

\subsection{Sobol indices}\label{subsec:Sobol example}

Now, we are interested in obtaining the first order Sobol indices of Ishigami Function presented in \verb|Equation| \ref{eq:16}. In order to obtain the Sobol index $So_i^\%$, we have to follow a simple process: To calculate $Var_{x_i}(E_{-x_i}[f(X)|x_i])$ we have to find $E[f(X)]$ fixing $x_i$ and varying the other parameters; after, we obtain the variance, by varying only $x_i$. When using the sampling method, it is necessary to choose a random value to $x_i$ then obtain the Expected Value of $f(X)$ by varying the others $x_{-i}$ parameters; this process must be repeated until we have the desired number of $E[f(X)]$, then, we extract the variance of these Expected Values \cite{Allard2009}.

For analytical solution, the process is more intuitive: we just need to obtain the $E_{-x_i}[f(X)|x_i]$ equation and then, obtain its variance over $x_i$ using \verb|Equations| \ref{eq:6}, \ref{eq:8}, \ref{eq:9} and \ref{eq:11}.

So, the Sobol percentage indices of \verb|Equation| \ref{eq:16} are calculated as follow:
\begin{equation} \label{eq:20}
Var_x(E_{-x}[Y|x]) = E_x[E_{-x}[Y|x]^2] - E_x[E_{-x}[Y|x]]^2
\end{equation}
with
\begin{eqnarray*}
E_{-x}[Y|x] &=& (\pi /10-(-\pi /10))^{-2}\int_{-\pi /10}^{\pi /10}\int_{-\pi /10}^{\pi /10}{sin(x)+7sin^2(y)+0.1z^4sin(x)}dydz \\
&=&25\pi^{-1}(0.394861sin(x)+0.0891373)  \\
\end{eqnarray*}
and
\begin{eqnarray*}
E_x[E_{-x}[Y|x]^2] &=& (\pi /10 -(-\pi /10))^{-1}\int_{-\pi /10}^{\pi/10}{(25\pi^{-1}(0.394861sin(x)+0.0891373))^2}dx  \\
&=&0.0832479\\
\end{eqnarray*}
\begin{eqnarray*}
E_x[E_{-x}[Y|x]]^2 &=& \left((\pi /10 -(-\pi /10))^{-1}\int_{-\pi /10}^{\pi/10}{25\pi^{-1}(0.394861sin(x)+0.0891373)}dx \right)^2 \\
&=&0.05098\\
\end{eqnarray*}
So, $Var_x(E_{-x}[Y|x])=(0.0832479-0.05098)=0.0322679$. Calculating the other parameters, we have:
\begin{equation} \label{eq:21}
Var_y(E_{-y}[Y|y]) = E_y[E_{-y}[Y|y]^2] - E_y[E_{-y}[Y|y]]^2
\end{equation}
\begin{eqnarray*}
E_{-y}[Y|y] &=& (\pi /10-(-\pi /10))^{-2}\int_{-\pi /10}^{\pi /10}\int_{-\pi /10}^{\pi /10}{sin(x)+7sin^2(y)+0.1z^4sin(x)}dxdz \\
&=&7sin^2(y)  \\
\end{eqnarray*}
\begin{eqnarray*}
E_y[E_{-y}[Y|y]^2] &=& (\pi /10 -(-\pi /10))^{-1}\int_{-\pi /10}^{\pi/10}{(7sin^2(y))^2}dy  \\
&=&0.0910763\\
\end{eqnarray*}
\begin{eqnarray*}
E_y[E_{-y}[Y|y]]^2 &=& \left((\pi /10 -(-\pi /10))^{-1}\int_{-\pi /10}^{\pi/10}{7Sin^2(y)}dx \right)^2 \\
&=&0.05098\\
\end{eqnarray*}
and finally for $z$, we have:
\begin{equation} \label{eq:22}
Var_z(E_{-z}[Y|z]) = E_z[E_{-z}[Y|z]^2] - E_z[E_{-z}[Y|z]]^2
\end{equation}
\begin{eqnarray*}
E_{-z}[Y|z] &=& (\pi /10-(-\pi /10))^{-2}\int_{-\pi /10}^{\pi /10}\int_{-\pi /10}^{\pi /10}{sin(x)+7sin^2(y)+0.1z^4sin(x)}dxdy \\
&=&\frac{7(4\pi - 5\sqrt{10-2\sqrt{5}})}{8\pi}  \\
\end{eqnarray*}
\begin{eqnarray*}
E_z[E_{-z}[Y|z]^2] &=& (\pi /10 -(-\pi /10))^{-1}\int_{-\pi /10}^{\pi/10}{\left(\frac{7(4\pi - 5\sqrt{10-2\sqrt{5}})}{8\pi}\right)^2}dz  \\
&=&0.05098\\
\end{eqnarray*}
\begin{eqnarray*}
E_z[E_{-z}[Y|z]]^2 &=& \left((\pi /10 -(-\pi /10))^{-1}\int_{-\pi/10}^{\pi /10}\frac{7(4\pi - 5\sqrt{10-2\sqrt{5}})}{8\pi}dz\right)^2 \\
&=&0.05098\\
\end{eqnarray*}

In summary, the $Var_y(E_{-y}[Y|y])=(0.0910763-0.05098)=0.0400963$ and  $Var_z(E_{-z}[Y|z])=(0.05098-0.05098)=0$.

\subsection{Example Results}\label{subsec:results}

Results obtained from Sections \ref{subsec:Example Variance} and \ref{subsec:Sobol example} are similar and may be used to demonstrate the influences of parameters $x,y,z$. \verb|Table| \ref{tab:1} depicts a summary of examples presented and compare these results with those extracted from \cite{Allard2009}.

\begin{table}[h]
\caption{\label{tab:1}Results of Examples}
\centering
\begin{tabular}{llllllp{7.4cm}}
\hline
Indices & Analytical Var &  Var Indices\cite{Allard2009} & Analitycal Sobol & Sobol R\cite{Allard2009}  & Sobol MatLab\cite{Allard2009}  \\\hline
$S_x^\%$  & $44.58\%$ & $44.8\%$ & $44.59\%$ & $41.55\%$ & $41.37\%$ \\
$S_y^\%$  & $55.42\%$ & $55.2\%$ & $55.41\%$ & $58.46\%$ & $58.62\%$ \\
 $S_z^\%$ & $0\%$ & $0\%$ & $0\%$ & $0\%$ & $0\%$\\ \hline
\end{tabular}
\end{table}

As it's possible to see in \verb|Table| \ref{tab:1}, the results for variance and Sobol indices analytically computed have a difference of just $0.01\%$. This difference can be explained by the precision of integrations values. Comparing with results presented on \cite{Allard2009} we can see the differences between the analytical solutions and the sampling-based solutions, mainly if we compare the Sobol sampling-based results. These differences can be explained by the nature of the analysis, since, the analysis presented in \cite{Allard2009} has been performed by sampling methods, that by nature have a random error.

\section{AVaN Pack} \label{sec:JSAvan}

In this section, we present the main features of AVaN Pack\footnote{AVaN Pack is a GPL-3.0 licensed software and is available in \url{https://github.com/dr-eduardovasconcelos/AVaNPack}}. This package is composed by two scripts (Variance Script and Sobol Script) that compute the sensitivity contribution of a function through a numerical solution. We decided to split the contribution of this manuscript into two programs to increase the code readability, since, these solutions are complex. Variance and Sobol scripts are small programs developed in Javascript (JS). These programs can find the individual percentage contributions of parameters of one equation. We have chosen JS to code these programs for some reasons:

\begin{itemize}
  \item JS is possibly the most used programming language throughout the world;
  \item The pure and old JS code can be executed practically in every smart device, provided it supports a web browser;
  \item Users don't need of any further structure such as servers or frameworks to execute a JS code, only the text source file;
  \item JS is a flexible language and codes can easily be modified and extended;
  \item With JS eval function, it's possible to convert symbolic equations into JS native functions;
\end{itemize}

So, two single-file codes have been developed to compute variance-based sensitivity analysis of any solvable and integrable equation. To keep these programs as simple as possible, we do not add any JS library or framework such as JQuery or Angular in our programs. With this, we can ensure that this program can execute even in old browsers, once they give support to JSON objects.

Following, we present the main features of each program.

\subsection{Variance Script}\label{subsec:js avan}

\begin{figure}[h]
\centering
\includegraphics[scale=0.6]{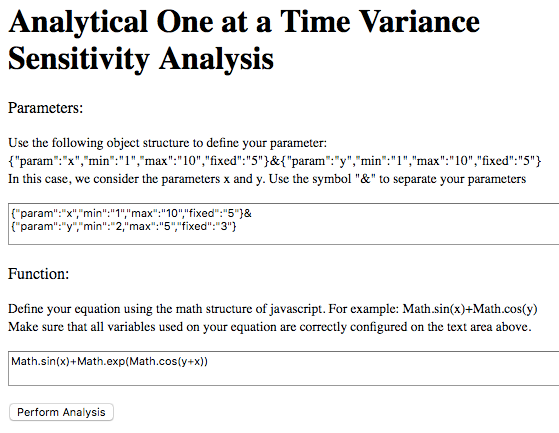}
\caption{\label{fig:1} Graphical User Interface of Variance Script}
\end{figure}

The variance script has a simple user interface composed by two HTML text areas and one button as showed in \verb|Figure| \ref{fig:1}. The first text area is used for parameter descriptions, the second one is used for the equation input. Parameters must be defined in JSON format. This decision allowed us to keep the program as simple as possible, decreasing the number of HTML elements and codes that should be necessary to treat these elements. Parameters inserted in the first text area must be described following 4 JSON parameters: 

\begin{itemize}
    \item "param" - represents the name of the parameter; 
    \item "min" - contains the minimum value assumed for this parameter;
    \item "max" - contains the maximum value;
    \item "fixed" - The fixed value for these parameters.
\end{itemize} 

For equation $ sin (x) + cos (y) $ the text inserted into the first text area must be: $ \{ "param": "x", "min": "1", "max": "10", "fixed": "5" \} \& \{"param": "y", "min": "2", "max": "5", "fixed": "3" \} $. Note that we have used "\&" to separate each parameter.
The equation must be inputted into the second text area using the JS math format. In this case, users must write their equations as if they were programming in Javascript. For example, let's suppose that an user wants to analyze equation $sin(x)+e^{cos(y+x)}$. Hence, the text that must be inputted into the second text area is: $Math.sin(x)+Math.exp(Math.cos(y+x))$.

To solve the analytical variance, the variance script makes use of numerical integration to find the expected values. The JS integration algorithm used is presented as follows:

\begin{algorithm}[H]
\SetAlgoLined
\caption{Numerical Integration}
\SetKwFunction{function}{function numericalIntegration}
\function{$f, a, b$} \{\\
    \quad \textbf{var} $delta = 0.00001*(b-a)$;\\
    \quad \textbf{var} $area = 0$;\\
    \quad \textbf{for} (\textbf{var} $i = a; i < b - delta; i += delta$) \{\\
        \quad \quad $area += (delta) * ((f(i) + f(i + delta)) / 2);$\\
    \quad \}\\
    \quad \textbf{return} $area;$\\
\}\\
\end{algorithm}
\ \\
Function $numericalIntegration()$ uses the trapezoidal method of integration to find the area of a function between $a$ and $b$. This method of integration consists in multiplying the height $(b-a)$ with the average of two bases $f(a)$ and $f(b)$. The variable \emph{delta} stores the partial heights of each trapezium that composes the integration. The value $0.00001$ represents the granularity of the process and, with this value, we guarantee the existence of $100000$ trapeziums per integration. We assume that the numerical integration imposes an error to results, but, we argue that different from the sampling-based solution, the error produced is not random, actually, this is a precision error.

With the method of integration defined, it is necessary to define the method to compose the JS functions. As we can see in \emph{numericalIntegration()} function, parameter \emph{f} is a single-parameter function, and contains the equation passed by the user. But, as a sensitivity analysis is made over an equation that has a set of parameters, define one single parameter function is unfeasible. So, to solve this problem, we used the flexibility of JS to create a set of texts that contain the partial functions for integration. After, we make use of code \emph{eval()} in order to turn these partial functions into JS functions. The definition of these function has been split into two phases: in the first, we need to create a multi-parameter function to execute the user equation; in the second phase, we define single parameter functions that will be used for numerical integration. The code used to create the first function is showed as following:

\begin{algorithm}[H]
\SetAlgoLined
\caption{String Equation Contruction}

\textbf{var} $equationInJSFmt = "function\ equation("$;\\
\textbf{for}(\textbf{var} $i = 0; i < parameters.length; i++$)\{ \\
                        
    \quad $equationInJSFmt += parameter[i].param;$\\
                        
     \quad $equationInJSFmt += (i==parameters.length-1)\ ?\ ")\{ return\ "+equationString+";\}"\ :\ ","$;
    
\}

$eval(equationInJSFmt);$

\end{algorithm}

The variable \emph{parameters} contains an array of JS objects, each of these objects contains the parameters defined into the first text area (See \verb|Figure|\ref{fig:1}). Variable \emph{equationString} contains the equation defined into the second text area. At the end of the definition of \emph{equation()}, we will have one function such as "\emph{function equation(x,y)\{returns Math. sin(x) + Math.cos(y);\};"} for the equation \emph{Math.sin(x)+Math.cos(y)}. After all, the function \emph{eval()} converts the text into a JS function.

To find the variance of $f(x)$ we need to calculate $f(x)^2$, thus we have to create a second function to be used to compute $E[f(x)^2]$. This second function is defined as the following:

\begin{algorithm}[H]
\SetAlgoLined
\caption{Creating Equation Moment}

\textbf{var} $momentEquationString = equationInJSFmt.replace("equation",
"equationMoment")$
$.replace("return", "return Math.pow(")$
$.replace(";", ",2);");$\\
$eval(momentEquationString);$

\end{algorithm}

This code basically uses the replace function to change \emph{equation} into \emph{equationMoment}. As a result, the String \emph{momentEquationString} will contain "\emph{function equationMoment(x,y) \{return Math.pow(Math.sin(x)+Math.cos(y),2);\};}". 

Functions \emph{equation} and \emph{equationMoment} are the first equations used in our solution. Now, we have to define the functions that will be directly used by \emph{numericalIntegration()}. So the definition of these single-parameter functions is defined as follow:

\begin{algorithm}[H]
\SetAlgoLined
\caption{Definition of \_fun and \_funMoment}

\textbf{var} $\_fun;$ \\
\textbf{var} $\_funMoment;$ \\
    
\textbf{var} $transientFunctionEqString = "\_fun = function(\_param)\{return \  
        equation(";$\\

\textbf{for} (\textbf{var} $j = 0; j < parameters.length; j++$) \{\\
\quad $transientFunctionEqString += (j==i)\ ?\ " \_param"\ :\ parameters[j].fixed + "";$\\
\quad $transientFunctionEqString += (j != parameters.length - 1)\ ?\ ","\ :\ ");\}";$
    \\
\}

$transientFunctionMoString = transientFunctionEqString$
     $.replace("equation", "equationMoment")$
     $.replace("\_fun", "\_funMoment");$\\

$eval(transientFunctionEqString);$\\
$eval(transientFunctionMoString);$

\end{algorithm}

This code repeats for each parameter on analysis. Functions \emph{\_fun} and \emph{\_funMoment} contain JS native functions and will be used to compute the expected values $E[f(x)]^2$ and $E[f(x)^2]$. The idea of the above code is create a function that receives one parameter and changes the others for their respective fixed values. For the function \emph{function equation(x,y)\{...} the resultant function for the first iteration is "\emph{\_fun = function(\_param)\{return\  equation(\_param, 3);\};}" if the fixed value of \emph{y} is 3.

Once the functions have been defined the rest of the process consists in integrate each function using the following code:

\begin{algorithm}[H]
\SetAlgoLined
\caption{Calculation of Variance Indices}

\textbf{var} $variances = new\  Array(parameters.length);$\\

\textbf{for}(\textbf{var} $i = 0; i < parameters.length; i++$)\{ \\
    
   \quad \textbf{var} $min = parseFloat(parameters[i].min);$ \\
   \quad \textbf{var} $max = parseFloat(parameters[i].max);$\\
    
    \quad \textbf{var} $squareEsperance = numericalIntegration(\_fun, min, max) / (max - min);$\\
                     
    \quad $squareEsperance = Math.pow(squareEsperance, 2);$\\
                       
    \quad \textbf{var} $moment = numericalIntegration(\_funMoment, min, max) / (max - min);$\\

    \quad $variances[i] = moment - squareEsperance;$
    
\}

\end{algorithm}

After this process, we obtain the analytical variances of parameters, and by using the \verb|Equation| \ref{eq:15}, we can obtain the first order percentage influences.

\subsection{Sobol Script} \label{subsec:savan}

The user interface of Sobol script is similar to that showed in \verb|Figure| \ref{fig:1}.

As Sobol analysis is different from variance analysis, the Sobol Script was designed to accommodate these differences. First, in Sobol analysis, it is necessary to perform multiple variable integrations, which increases the overall complexity of any solution. Second, to solve these integrations numerically it's necessary to use multiple loops, what increase the computational complexity. With this in mind, the first step necessary to compute the Sobol indices is to calculate the value of $E_{-x_i}[f(X)|x_i]$. This expected value is computed by function \emph{nMinusXiAverageIntegration()} that is showed as follow: 

\begin{algorithm}[H]
\SetAlgoLined
\caption{First part of nMinusXiAverageIntegration function}
\SetKwFunction{function}{function nMinusXiAverageIntegration}

\function{equation, xi, X, delta\_base} \{ \\
    \quad \textbf{var} $delta\_X = new Array(X.length);$\\
    \quad \textbf{var} $delta\_inc_X = new Array(X.length);$ \\
    \quad \textbf{var} $baseProduct = 1.0;$\\
    \quad \textbf{var} $aMinusbProd = 1.0;$\\

    \quad \textbf{for} (\textbf{var} $i = 0; i < X.length; i++$) \{\\
        \quad \quad $delta\_X[i] = delta\_base * (X[i].max - X[i].min);$\\
        \quad \quad $delta\_inc\_X[i] = X[i].min;$\\

        \quad \quad $baseProduct *= delta\_X[i];$\\
        \quad \quad $aMinusbProd *= X[i].max - X[i].min;$ \\
        \quad \quad $delta\_X[i] = delta\_base * (X[i].max - X[i].min);$\\
        \quad \quad $delta\_inc\_X[i] = X[i].min;$
        .
        .
        .

\end{algorithm}

Function \emph{nMinusXiAverageIntegration()} receives 4 arguments: The first is a literal that contains the equation that must be analyzed; the second is an object that receives the name and the value of $x_i$; \emph{X} is an array of objects representing the other parameters of equation, and \emph{delta\_base} is a float number that represents the granularity of our integration. The variable \emph{delta\_X} is an array that contains the heights of parameters of the array \emph{X} and \emph{delta\_inc\_X} is an array of numbers used for controlling the process of integration.

The following code shows the process of defining the literal functions used on analyzes.

\begin{algorithm}[H]
\SetAlgoLined
\caption{Second part of nMinusXiAverageIntegration function}

\textbf{var} $transBaseFun = "\_baseFun = function (" + xi.param + ","$;\\
                
\textbf{var} $\_basefun$;\\

\textbf{for} (\textbf{var} $i = 0; i < X.length; i++$) \{ \\

    \quad $transBaseFun = transBaseFun + X[i].param+((i != X.length - 1)\ ?\ "," \ : \ ")\{return\ " +
            equation + ";\};");$\\
  
\} \\     
$eval(transBaseFun);$
        .
        .
        .
\end{algorithm}

Variable \emph{transBaseFun} has the base function of the integration, which computes the equation passed as parameters. The first argument of \emph{\_basefun} is $x_i$, the other arguments are inserted into the function according to the sequence of array \emph{X}. If the equation passed as argument is $x+y+z$ then the resultant funtion is "\emph{\_basefun(y,x,z)\{return x+y+x;\};}". Note that, in this case, the variable \emph{y} is the $x_i$ in this process.

Function \emph{\_basefun} contains only the analyzed equation. To make the multiple integrations, we need another function that makes the sums necessary for the trapezoidal integration. This second function is assembled as follow:

\begin{algorithm} [H]
\SetAlgoLined
\caption{Third part of nMinusXiAverageIntegration function}
\textbf{var} $transIntFun = "\_intFun = function(delta\_inc\_X)\{
            return\ \_baseFun("     + xi.value + ",";$\\
\textbf{var} $\_intFun;$\\
      
\textbf{for} (\textbf{var} $i = 0; i < Math.pow(2, X.length); i++$) \{\\
    \quad \textbf{for} (\textbf{var} $j = 0; j < X.length; j++$) \{\\

       \quad \quad  $transIntFun += "delta\_inc\_X[" + j + "]";$ \\

        \quad \quad \textbf{if} $(i \& Math.pow(2, j)) != 0)$ 
           \{\ \ $transIntFun += "\ +\ " + delta\_X[j];\ \ \}$\\
        
\quad \quad $transIntFun += (j == X.length - 1)\ ? \ ")\ "\ :\ ", ";$
       
    \quad \}\\

\quad $transIntFun += (i == Math.pow(2, X.length) - 1)\ ?\ ";\};"\ : \ "+ \_baseFun(" + xi.value + ",";$

\} \\
        
$eval(transIntFun);$
        .
        .
        .
\end{algorithm}

The base of function \emph{\_intFun} represents the generalization of trapezoidal method for integration of multiple variables. To explain the resultant format of \emph{\_intFun}, let's consider that \emph{\_basefun} contains 3 arguments x, y and z, and y is the analyzed parameter. Let's consider also that the fixed value of y is 3 and the other parameters are being integrated with the intervals a, b and c, d. So, the general form of this trapezoidal integration is "\emph{(b-a)*(d-c) * (\_basefun(3,a,c)+\_basefun(3,a+b,c)+\_basefun(3,a,c+d)+ \_basefun(3,a+b,c+d))/4}". Function \emph{intFun} computes the sums of \emph{\_basefun} and the product among \emph{(b-a)} and \emph{(d-c)} is represented by the variable \emph{baseProduct} that is introduced into the next piece of code. Basically, in function \emph{\_intFun} the intervals of each parameter are given by array \emph{delta\_X}.

Once we have the two functions \emph{\_basefun} and \emph{\_intFun}, the next step is to compute the expected value $E_{-x_i}[f(X)|x_i]$ through the following piece of code:

\begin{algorithm} [H]
\SetAlgoLined
\caption{Last part of nMinusXiAverageIntegration function}       
\textbf{var} $partialArea = 0.0;$\\
        
\textbf{for} (\textbf{var} $i = 0; i < Math.pow(1.0 / delta\_base, X.length); i++$) \{ \\

   \quad $partialArea += baseProduct * (\_intFun(delta\_inc\_X)
                / Math.pow(2,X.length));$\\

   \quad  $delta\_inc\_X[0] = (delta\_inc\_X[0] + delta\_X[0]);$\\

    \quad \textbf{if} ($delta\_inc\_X[0] > X[0].max-delta\_X[0]$) \{ \\
       \quad \quad  $delta\_inc\_X[0] = X[0].min;$ \\

       \quad \quad  \textbf{for} (\textbf{var} $j = 1; j < X.length; j++$) \{ \\
          \quad \quad \quad  \textbf{if} ($delta\_inc\_X[j - 1] == X[j - 1].min$) \{ \\
             \quad \quad \quad \quad    $delta\_inc\_X[j] = (delta\_inc\_X[j] + delta\_X[j]);$ \\
             \quad \quad \quad \quad \textbf{if} ($delta\_inc\_X[j] > X[j].max-delta\_X[j]$)\{\ \  
               $delta\_inc\_X[j] = X[j].min;$\ \ \} \\
              
            \quad \quad \quad \} \\
        \quad \quad \}\\

    \quad \} \\

\} \\

\textbf{return} $partialArea / aMinusbProd;$\\

\end{algorithm}

Since for computing an n-variable numerical integration we need of n loops, in Sobol Script, to adapt the algorithm for any number of parameters, we have decided to create a bigger single loop. This loop will have $(1/delta\_base)^n$ iterations, that is the same complexity of having n loops. After calculating the sum of trapeziums, we need to update the values of \emph{delta\_inc\_X} array. As \emph{delta\_inc\_X[i]} has a value between \emph{X[i].min} and \emph{X[i].max}, this value must be incremented until it reaches the maximum value of parameter \emph{X[i]}. Once \emph{delta\_inc\_X[i]} reaches its maximum value, the program resets it and increment \emph{delta\_inc\_X[i+1]}. So, in the above piece of code, \emph{delta\_inc\_X[0]} is incremented in every iteration, and always which this variable reaches its maximum value, the code scrolls through the array to update the other parameters. Then, after computing the main sum, the function \emph{nMinusXiAverageIntegration} returns the ratio between the sum result and the product of maximum and minimum values of parameters, this ratio represents the multi-variable expected value.

Once we have computed $E_{-x_i}[f(X)|x_i]$ it's necessary to compute $Var_{x_i}(E_{-x_i}[f(X)|x_i])$. The process to compute $Var_{x_i}$ for symbolic integration is known, we need to calculate the resultant equation $E_{-x_i}[f(X)|x_i]$ and then integrate it to obtain the variance. However, for numerical integration it cannot be done, since it's not possible to integrate a multi-variable function in parts. Then, the following piece of code shows the solution.

\begin{algorithm}[H]
\SetAlgoLined
\caption{Sobol Indices Calculation}   

\textbf{var} $delta\_base = 0.001;$\\

\textbf{for}(\textbf{var} $i = 0;i<parameters.length;i++$)\{ \\
   \quad \textbf{var} $minVal = parseFloat(parameters[i].min);$\\
    \quad \textbf{var} $maxVal = parseFloat(parameters[i].max);$\\

    \quad \textbf{var} $delta\_xi = (maxVal - MinVal)*delta\_base;$

    \quad \textbf{var} $areaMomento = 0.0;$\\
    \quad \textbf{var} $areaSquare = 0.0;$\\

    \quad \textbf{for} (\textbf{var} $j = minVal; j < maxVal; j += delta\_xi$) \{ \\
                        
       \quad \quad $xi.value = j + delta\_xi / 2.0;$\\
        \quad \quad \textbf{var} $partialInteg = nMinusXiAverageIntegration(equationString, xi, X, delta\_base);$\\
        \quad \quad $areaMomento += delta\_xi * Math.pow(partialInteg, 2);$\\
        \quad \quad $areaSquare += delta\_xi * partialInteg;$\\

    \quad \} \\
    
    \quad $V[i] = areaMomento / (maxVal - minVal) -
            Math.pow(areaSquare / (maxVal - minVal), 2);$\\
    \quad $sumV += V[i];$\\

\}
\end{algorithm}

In this solution, we basically scroll through the minimum and maximum values in \emph{delta\_xi} steps, compute the expected value of $-x_i$ parameters and then, sum these expected values to obtain $E_{x_i}[E_{-x_i}[f(x)|x_i]^2]$ and $(E_{x_i}[E_{-x_i}[f(x)|x_i]])^2$. Note that the value of $x_i$ is defined as \emph{j + delta\_xi/2.0}, it's because, for this integration, we have used the rectangle rule. This measure allowed us to reduce the complexity of our solution, since, to compute the trapezoidal rule we need to call \emph{nMinusXiAverageIntegration()} twice for each value of $x_i$. The code "\emph{areaMomento += delta\_xi * Math.pow(partialInteg, 2)}" represents the partial calculation of $E_{x_i}[E_{-x_i}[f(x)|x_i]^2]$ as presented in \verb|Equation| \ref{eq:9}.
Finally, all variances are stored into array \emph{V}. From \emph{V}, it's possible to compute the Sobol percentage indices of user equation. The complexity of Sobol script is $O(n)=n\times(delta\_base^{-n})$ for $delta\_base<1$ and $n$ the number of parameters.

\subsection{Experiments}\label{subsec:experiments}

The objective of this section is to compare the outcomes of Variance and Sobol scripts with results presented in \verb|Table| \ref{tab:1}. The object of analysis is the Ishigami Function presented on \verb|Equation| \ref{eq:16}. The JS format of Ishigami Function is presented as following:
$Math.sin(x) + 7*Math.pow(Math.sin(y),2) + 0.1*(Math.pow(z,4))*Math.sin(x)$.

The above text is set into the second text area of both programs. The parameters in JSON format are described as following:

Variance Script\\
\{"param":"x","min":"-0.31415926535897932384626433",
"max":"0.31415926535897932384626433","fixed":"0.0"\} \&\\ 
\{"param":"y","min":"-0.31415926535897932384626433",
"max":"0.31415926535897932384626433","fixed":"0.0"\} \&\\
\{"param":"z","min":"-0.31415926535897932384626433",
"max":"0.31415926535897932384626433","fixed":"0.0"\}

Sobol Script\\
\{"param":"x","min":"-0.31415926535897932384626433",
"max":"0.31415926535897932384626433"\} \&\\ 
\{"param":"y","min":"-0.31415926535897932384626433",
"max":"0.31415926535897932384626433"\} \&\\
\{"param":"z","min":"-0.31415926535897932384626433",
"max":"0.31415926535897932384626433"\}

The values of \emph{delta} and delta\_base for both Variance Script and Sobol Script, has been set as \emph{0.00001} and \emph{0.001}, respectively. The reason for setting \emph{delta\_base} with a greater value then \emph{delta} is the complexity of Sobol Script code. \verb|Table| \ref{tab:2} shows a summary of experiments. 
As it's possible to observe, both programs have produced exactly the same results when compared with the mathematical results. It's important to mention that the precision error imposed by the numerical integration can't be observed in results presented in \verb|Table| \ref{tab:2} because of the decimal places of percentage format. Note that the fact of both programs produced results similar to the mathematical solution for this particular values of \emph{delta} and \emph{delta\_base} doesn't mean that these results will repeat for any experiment performed on AVaN Pack. So, whenever questions remain about the outcomes, it will be necessary to decrease the values of \emph{delta} or \emph{delta\_base}.

\begin{table}[h]
\caption{\label{tab:2}Program outcomes}
\centering
\begin{tabular}{lllllp{7.4cm}}
\hline
Indices & Variance Script & Analytical Var & Sobol Script & Analitycal Sobol \\\hline
$S_x^\%$  & $44.58\%$ & $44.58\%$ & $44.59\%$ & $44.59\%$\\
$S_y^\%$  & $55.42\%$ & $55.42\%$ & $55.41\%$ & $55.41\%$ \\
 $S_z^\%$ & $0\%$ & $0\%$ & $0\%$ & $0\%$\\ \hline
\end{tabular}
\end{table}

\bibliographystyle{unsrt}  

\bibliography{AvanPack}  %%% Remove comment to use the external .bib file (using bibtex).

%\bibliography{references}

\end{document}